\documentclass{aastex}          %% The manuscript based on AASTeX v5.x
\usepackage{spr-astr-addons}    %% mimicing ASTR journal style
\usepackage{natbib}
\usepackage{graphicx}
\usepackage{bm}
\usepackage[breaklinks=true,colorlinks=true,linkcolor=blue,urlcolor=blue,citecolor=blue]{hyperref}

\def\dg{\color{blue}}
\begin{document}

\title{Theory of neutrino emission from nucleon-hyperon matter in neutron stars:
       Angular integrals}

\shorttitle{Neutrino emission from nucleon-hyperon matter}

\shortauthors{Kaminker et al.}

%% Author and Affilations
\author{A. D. Kaminker\altaffilmark{1}}
\and
\author{D. G. Yakovlev\altaffilmark{1}}
\and
\author{P. Haensel\altaffilmark{2}}
%\affil{1}{Ioffe Physical Technical Institute, Politekhnicheskaya
%26, 194021, St Petersburg, Russia}
\email{kam.asro@mail.ioffe.ru} %% non-output

%% Alternate Affilations
\altaffiltext{1}{Ioffe Physical Technical Institute, Politekhnicheskaya
26, 194021, St Petersburg, Russia}
\altaffiltext{2}{N.\ Copernicus Astronomical Center, Bartycka 18, 00-716 Warsaw, Poland}
%\altaffiltext{3}{}

\begin{abstract}
Investigations of thermal evolution of neutron stars with hyperon
cores require neutrino emissivities for many neutrino reactions
involving strongly degenerate particles (nucleons, hyperons,
elect\-rons, muons). We calculate the angular integrals $I_n$ (over
orientations of momenta of $n$ degene\-rate particles) for major
neutrino reactions with $n=$3, 4, 5 at all possible combinations of
particle Fermi momenta. The integrals $I_n$ are necessary
ingredients for constructing a uniform database of neutrino
emissivities in dense nucleon-hyperon matter. The results can also
be used in many problems of physical kinetics of strongly degenerate
systems.
\end{abstract}

%\pacs{05.30.Fk, 71.18.+y, 42.10}

%\vspace{2pc} \noindent{\it Keywords}:

\section{Introduction}
\label{s:introduc}

It is well known that thermal evolution of not too cold neutron
stars is regulated by the neutrino emission from superdense matter
in neutron star cores. In order to model the thermal evolution one
needs the emissivies of many neutrino reactions which can operate
and produce an efficient neutrino cooling of these stars (e.g.,
\citealt{YKGH01}). Consider, for instance, neutron star cores, which
are massive and bulky internal regions of neutron stars
\citep{ST83}. They are thought to contain uniform nuclear liquid
 of density $\rho$ ranged from $\sim \rho_0/2$ to $\sim10-20$
$\rho_0$, where $\rho_0 \approx 2.8 \times 10^{14}$ g~cm$^{-3}$ is
the density of standard nuclear matter at saturation. A neutron star
core can be divided into the outer core ($\rho \lesssim 2 \rho_0$)
composed of neutrons (n) with some admixture of protons (p),
electrons (e) and muons ($\mu$), and the inner core ($\rho \gtrsim 2
\rho_0$) containing the same particles and possibly other ones (for
instance, hyperons). All constituents of the matter (n, p, e, $\mu$,
hyperons) are strongly degenerate fermions. These particles can
participate in many reactions producing neutrinos.  Because in half
a minute after their birth neutron stars become fully transparent
for neutrinos, the neutrinos immediately escape from the star and
cool it.

Schematically, the neutrino emissivity [erg~cm$^{-3}$~s$^{-1}$] for
any reac\-tion can be written as
\begin{equation}
   Q= (2\pi)^4  \int {\rm d}\Gamma \;{\cal M}_{\cal F I}\,\epsilon_\nu \,
\delta(\bm{P}_{\cal F}-\bm{P}_{\cal I})
   \;\delta(E_{\cal F}-E_{\cal I})\,F_{\cal F I}.
\label{e:R}
\end{equation}
Here, $\epsilon_\nu$ is the energy of generated neutrino (or neutrinos), ${\cal I}$
and  ${\cal F}$ label initial and final states of a system, while
$i$ and $f$ label corresponding states of reacting particles;
$\bm{P}_{\cal I}=\sum_i \bm{p}_i$ and $\bm{P}_{\cal
F}=\sum_f \bm{p}_f$ denote, respectively, total momenta of reacting
particles in the ${\cal I}$ and ${\cal F}$ states, with $\bm{p}$
being a one-particle momentum; $E_{\cal I}=\sum_i \epsilon_i$ and
$E_{\cal F}=\sum_f \epsilon_f$ are total energies of the particles
($\epsilon$ is a one-particle energy). The delta functions take into
account momentum and energy conservation in a reaction event. The
factor $F_{\cal  F I}$ is
\begin{eqnarray}
         F_{\cal FI} &=&\left(\prod_{i}f_i\right)\,\left( \prod_f (1-f_f)
     \right),\nonumber \\
        f_i& =& \left[\exp\left(\frac{\epsilon_i-\mu_i}{k_{\rm B}T}
     \right)+1 \right]^{-1}.
\label{e:F}
\end{eqnarray}
It contains the product of Fermi-Dirac functions $f_i$ for particles
in the initial states  and the product of blocking functions
$(1-f_f)$ for particles in the final states; $\mu_i$ is the chemical
potential, $T$ the temperature, and $k_{\rm B}$ the Boltzmann
constant. In what follows, we take into account that neutron star
matter is fully transparent for neutrinos (e.g., \citealt{YKGH01}).
Then the chemical potential of neutrinos is zero, $\mu_\nu=0$, and
the approximation of massless neutrinos is excellent. Other initial
or final reacting fermions $j$ (which belong to the dense matter)
are assumed to be strongly degenerate particles of any relativity;
their energies $\epsilon_j$ and chemical potentials $\mu_j$ may
include or exclude the rest-mass energy, $m_jc^2$. The quantity
${\cal M}_{\cal F I}$ in Eq.\ (\ref{e:R}) is proportional to the
squared matrix element for a given reaction summed over spin states.
Finally,
\begin{equation}
   {\rm d}\Gamma = \prod_{l}\, \frac{{\rm d}\bm{p}_l}{(2 \pi \hbar)^3}
\label{e:G}
\end{equation}
is the product of densities of states of all reacting partic\-les~$l=j$ and $\nu$.

It is well known (e.g., \citealt{Ziman,BP07,ST83}) that calculation
of the emissivities (\ref{e:R}), reaction rates or related
quantities in a strongly degenerate matter is greatly simplified
because the main contribution into corresponding integrals comes
from narrow thermal energy widths $|\epsilon_j - \mu_j|\ll k_{\rm
B}T$. Accordingly, one can usually employ the so called
energy-momentum decomposition detailed, e.g., in
\citet{Ziman,BP07,ST83}. It consists in fixing lengths of all
momenta of strongly degenerate reacting fermions to the
corresponding Fermi momenta ($|\bm{p}_j|=p_{{\rm F}j}$) and values
of energies of these particles to the corresponding chemical
potentials ($\epsilon_j=\mu_j$) in all functions of $\bm{p}_j$ and
$\epsilon_j$ which vary smoothly within thermal energy widths in
local elements near respective Fermi surfaces. Then in Eq.\
(\ref{e:G}) one can set ${\rm d}\bm{p}_j=p_{{\rm F}j}m_j^*\,{\rm
d}\epsilon_j\,{\rm d}\Omega_j$, where $m_j^*$ is the Landau
effective mass of a fermion $j$ at the Fermi surface, and ${\rm
d}\Omega_j$ is a solid angle element in the direction of $\bm{p}_j$.
The integration over particle momenta in Eq.\ (\ref{e:R}) is then
decomposed into the integration over energies d$\epsilon_j$ and over
solid angles d$\Omega_j$.

In further calculations of the emisivity $Q$ one often approximates
${\cal M}_{\cal F I}$ by its value $\langle {\cal M}_{\cal F I}
\rangle$ averaged over orientations of particle momenta. Then the
emissivity becomes
\begin{equation}
    Q= I_\epsilon I_\Omega,
\label{e:R1}
\end{equation}
where
\begin{equation}
    I_\Omega= \int \delta(\bm{P}_{\cal F}-\bm{P}_{\cal I})\,
        \prod_j \,{\rm d}\Omega_j
\label{e:I}
\end{equation}
is the integral over orientations of all particle momenta placed on
respective Fermi surfaces, while $I_\epsilon$ contains all other
terms (including $\langle {\cal M}_{\cal F I} \rangle$) and
integration over particle energies.

In a neutron star, generated neutrinos have much lower energies and
momenta than the particles of the matter; it is quite sufficient to
neglect neutrino momenta in the momentum-conserving delta function
 in Eq. (\ref{e:I}) (e.g., \citealt{YKGH01}). Then the integration over orientations of
neutrino momentum is trivial (e.g., gives a factor of $4 \pi$ for an
emission of one neutrino) and will be supposed to be included in
$I_\epsilon$. Accordingly, the angular integration in $I_\Omega$ is
performed only over orientations of momenta of strongly degenerate
fermions $j$ of the matter. The number of these fermions will be
denoted by $n$, so that $j=1,\ldots,n$ in $I_\Omega$.

The case of strongly interacting fermions (nucleons and hyperons in
dense nuclear matter) deserves a comment. We assume that the system
is non-superfluid, i.e., it is a normal Fermi liquid (see, e.g,
\citealt{BP07,LP1980}). Then the one-particle states with well
defined energies and momenta refer actually to elementary
excitations, called Landau quasiparticles. In a strongly degenerate
Fermi liquid, quasiparticles form a dilute Fermi gas. Therefore,
their distribution in momentum space can be well approximated by the
Fermi-Dirac one. The Fermi momenta for quasiparticles coincide with
those for real particles. These properties  justify  the use of
Eq.~(\ref{e:F}). In what follows, by  particles in Fermi liquids of
nucleons and hyperons we will mean quasiparticles.

Systems of strongly degenerate particles are also important in many
branches of physics. In particular, we can mention solid state
physics (degenerate electrons in metals and semiconductors; e.g.,
\citealt{Ziman,Kittel}), Fermi-liquid systems \citep{BP07} as well
as nuclear physics (symmetric nuclear matter in
atomic nuclei).

It is our aim to consider the angular integrals $I_\Omega$ for
reactions involving different particle species with various Fermi
momenta. These integrals determine the area of a hypersurface in
3$n$-dimensional momentum space which contributes to a given
reaction. The advantage of the integrals $I_\Omega$ is that they are
indepen\-dent of specific interparticle interactions. They depend
only on the total number $n$ of reacting particles and on Fermi
momenta of these particles, $p_{{\rm F}j}\equiv p_j$
($j$=1,\ldots$n$). For simplicity, we drop the subscript F because
all momenta are assumed to be on the Fermi surfaces. The angular
integrals $I_\Omega$ appear in many problems of Fermi systems (e.g.,
\citealt{Ziman,BP07,ST83}). Some approaches for calculating them are
described, for instance by \citet{ST83}.

However, there are plenty of cases realized for different Fermi
momenta. Our aim is practical, to present $I_\Omega$ for all
possible cases at $n\leq 5$. In particular, these cases correspond
to major neutrino reactions in nucleon-hyperon matter of neutron
stars.

Section \ref{s:remarks} outlines a general method for calculating
$I_\Omega$. Sections \ref{s:n*nd3}, \ref{s:n=4} and \ref{s:n=5}
present the calculations for $n\leq 5$. Applications for neutrino
reactions are briefly discussed in Section \ref{s:applic}, and we
conclude in Section \ref{s:conclusion}.

%%%%%%%%%%%%%%%%%%%%%%%%%%%%%%%%%%%%%%%%%%%%%%%
\section{General remarks}
\label{s:remarks}
%%%%%%%%%%%%%%%%%%%%%%%%%%%%%%%%%%%%%%%%%%%%%%%

Let us formulate some general properties of the angular integrals
$I_\Omega$ as functions of $n$ Fermi momenta. Since we integrate
over all possible orientations of $\bm{p}_j$, any inversion
$\bm{p}_j \to -\bm{p}_j$ does not change $I_\Omega$. Therefore,
\begin{eqnarray}
    I_\Omega &= &I_\Omega^{(n)}(p_1,\ldots,p_n) \nonumber \\
       & =& \int \delta\left(\bm{p}_1 + \ldots + \bm{p}_n \right)\,
        \prod_{j=1}^n \,{\rm d}\Omega_j.
\label{e:I1}
\end{eqnarray}

This will be our starting expression. The lengths of Fermi momenta,
$p_1, \ldots,p_n$, are treated as given numbers; physical { nature}  of
particle species $j$ is of no importance. For the sake of
convenience, we enumerate the particles { in the order} of decreasing {
Fermi} momenta,
\begin{equation}
   p_1 \geq p_2 \geq p_3 \geq \ldots \geq p_n.
\label{e:order}
\end{equation}

Because the delta function in Eq.\ (\ref{e:I1}) describes momentum
conservation, any angular integral is non-zero ($I_\Omega>0$) if at
least
\begin{equation}
   p_1 \leq p_2+\ldots+p_n.
\label{e:pconserv}
\end{equation}

In addition to integrals (\ref{e:I1}) it is often convenient
to introduce similar auxiliary integrals
\begin{eqnarray}
    \widetilde{I}_\Omega& = &\widetilde{I}_\Omega^{(n)}(p_1,\ldots,p_n,q)
\nonumber \\
        &=&\int \delta\left(\bm{p}_1+\ldots+\bm{p}_n+\bm{q} \right)\,
        \prod_{j=1}^n \,{\rm d}\Omega_j,
\label{e:tI1}
\end{eqnarray}
where $\bm{q}$ is a constant vector of arbitrary length limited by
momentum conservation. Because of isotropy of momentum space,
$\widetilde{I}_\Omega$ depends only on { $q=|\bm{q}|$}  (but not on
{ the direction} of $\bm{q}$). Using the definition (\ref{e:I1}) one
can present Eq.~(\ref{e:tI1}) in the form
\begin{equation}
      \widetilde{I}_\Omega^{(n)}(p_1,\ldots,p_n,q)= {1 \over 4\pi}\,
                  I_\Omega^{(n+1)}(p_1, \ldots,p_n;q).
\label{e:tI1=I}
\end{equation}
It is easy to show that
\begin{eqnarray}
&& I_\Omega^{(n)}(p_1,\ldots,p_n) \nonumber \\
&& = \int {\rm d}\bm{q}\,
       \widetilde{I}_\Omega^{(m)}(p_1,\ldots,p_m,q)\,
            \widetilde{I}_\Omega^{(n-m)}(p_{m+1},\ldots,p_n,q)
\nonumber    \\
        && =
        {1 \over 4\pi}\, \int\ q^2\ {\rm d}q\, I_\Omega^{(m+1)}(p_1,\ldots,p_m;q)\,
\nonumber \\
   && ~~  \times  I_\Omega^{(n-m+1)}(p_{m+1},\ldots,p_n;q),
\label{e:I=tItI}
\end{eqnarray}
with $m<n$. This equality greatly simplifies calculati\-ons of
angular integrals because it allows one to consider the reacting
particles as two subsystems (1,\dots,\ $m$) and ($m+1,\ldots,\ n$).
Then one can take auxiliary integrals for these subsystems
separately, which is simpler than calculate $I_\Omega^{(n)}$
directly. In this case $\bm{q}$ is a momentum transfer from one
subsystem to the other. A { partition}  of particles into these
subsystems is arbitrary. The resulting angular integral
$I_\Omega^{(n)}$ is, of course, independent of specific partition.

Let us consider
$I_\Omega^{(n)}$ with $n\leq 5$.

%%%%%%%%%%%%%%%%%%%%%%%%%%%%%%%%%%%%%%%%%%%%%%%%

\section{Reactions involving $n$=2 and 3 fermions}
\label{s:n*nd3}

If $n=2$ we have $\bm{p}_1=-\bm{p}_2$ and Eq.\ (\ref{e:I1})
yields
\begin{equation}
     I_\Omega^{(2)}(p_1,p_2)= \frac{2\ (2 \pi)}{p_1 p_2}\,\delta(p_1-p_2).
\label{e:n=2}
\end{equation}
In the case of $n=3$ and  $p_1<p_2+p_3$ (standard triangle
condition) one obtains
\begin{equation}
 I_\Omega^{(3)}(p_1,p_2,p_3)= \frac{2 (2\pi)^2}{p_1 p_2 p_3}.
\label{e:n=3}
\end{equation}
It is the basic expression to be used for all angular integrals with
$n>3$.

\section{Reactions involving $n$=4 fermions}
\label{s:n=4}

{ In the case} of four reacting fermions one can use Eq.\ (\ref{e:I=tItI})
and { divide the system of four particles} into two subsystems, say (1,\ 2) and (3,\
4). Then using Eq.~(\ref{e:n=3}) one comes to a general expression
($p_1<p_2+p_3+p_4$)
\begin{eqnarray}
 && I_\Omega^{(4)}(p_1,p_2,p_3,p_4) \nonumber \\
&& =  {1 \over 4 \pi}
 \int\ q^2\, {\rm d}q \, I_\Omega^{(3)}(p_1, p_2; q)\,
        I_\Omega^{(3)}(p_3, p_4; q)
\nonumber \\
  &&  =  \frac{2 (2 \pi)^3}{p_1 p_2 p_3 p_4} \, Q^{(4)} , \quad
    Q^{(4)}=q_\mathrm{max}-q_\mathrm{min},
\label{e:n=4}
\end{eqnarray}
where
$q_\mathrm{max}=p_3+p_4$,  $q_\mathrm{min}=\mathrm{max}(p_1-p_2,~
    p_3-p_4)$;
$q_\mathrm{max}$ is the maximum momentum transferred between
subsystems (1,2) and (3,4); $q_\mathrm{min}$ is the minimum momentum
transfer between these subsystems.

%T1%
%%%%%%%%%%%%%%%%%%%%%%%%%%%%%%%%%%%%%%%%%%%%%%%%%%%%%%%%%%%%%%%%%%

\begin{table}
%\small
\caption{Particular cases of $Q^{(4)}$ for 4 fermions} \label{tabQ4}
\begin{tabular}{ll}
%\rule{0.5\textwidth}{0.4pt}\\
\hline \hline
 Conditions  &   $Q^{(4)}$\\
\hline
%\tableline\\
  $p_1=p_2>p_3>p_4$ & $2p_4$ \\
%\tableline\\
%2  &  $p_1>p_2=p_3>p_4$ & \hspace{-0.2cm}
%      $\begin{array}{ll}
    $p_2=p_3,~~p_1+p_4>2p_2$ & $2p_2+p_4-p_1$ \\ %& {\rm at~}p_1+p_4>2p_2  \\
    $p_2=p_3,~~p_1+p_4\leq  2p_2$ & $2p_4$ \\  %   & {\rm at~}   p_1+p_4\leq  2p_2
%    \end{array}  $ \\
%\tableline\\
 $p_1>p_2>p_3=p_4$  &   $p_2+2p_3-p_1$   \\
%\tableline\\
 $ p_1=p_2>p_3=p_4 $ &  $2p_3$ \\
%\tableline\\
 $ p_1=p_2=p_3>p_4 $ &  $2p_4$ \\
%\tableline\\
 $ p_1>p_2=p_3=p_4 $ &  $3p_4-p_1 $ \\
%\tableline\\
 $ p_1=p_2=p_3=p_4 $ & $2p_1$ \\
%\rule{0.5\textwidth}{0.4pt}
\hline \hline
\end{tabular}
%\end{indented}
\end{table}
%%%%%%%%%%%%%%%%%%%%%%%%%%%%%%%%%%%%%%%%%%%%%%%%%%%%%%

In the general case of
$p_1 \geq p_2 \geq p_3 \geq p_4$ we have
\begin{equation}
    Q^{(4)} =   \left\{
    \begin{array}{ll}
    p_2+p_3+p_4-p_1 & {\rm at~}p_1+p_4 \geq p_2+p_3;\\
    2 p_4 \qquad & {\rm at~}p_1+p_4 < p_2+p_3.
    \end{array}
    \right.
\label{e:4diff}
\end{equation}
This expression is further simplified in several particular cases
listed in Table~\ref{tabQ4} -- for
two equal fermion momenta out of four,  two
different pairs of equal momenta,  three equal
momenta, and for equal momenta of all four
fermions.

%%%%%%%%%%%%%%%%%%%%%%%%%%%%%%%%%%%%%%%%%%%%%%%%%%%%
\section{Reactions involving $n=5$ fermions}
\label{s:n=5}

In the case of five reacting fermions one also can use Eq.\
(\ref{e:I=tItI})  and divide fermions into two subsystems, for
instance, (1,2) and (3,4,5), with $p_1\leq  p_2+p_3+p_4+p_5$ and
$p_1 \geq p_2 \geq p_3 \geq p_4 \geq p_5$. In this way we obtain
%%%%%%%%%%%%%%%%%%%%
\begin{eqnarray}
&&I_\Omega^{(5)}(p_1,p_2,p_3,p_4,p_5)
\nonumber \\
&&={1 \over 4\pi}\ \int q^2 \, {\rm
d}q\, I_\Omega^{(3)}(p_1, p_2; q)\,
        I_\Omega^{(4)}(p_3, p_4, p_5; q).
\label{e:I5=I3I4}
\end{eqnarray}
%%%%%%%%%%%%%%%%%%%%%%%%%%%%%%%%%%%

\begin{figure*}[htb!]
\centering
\includegraphics[width=0.90\textwidth]{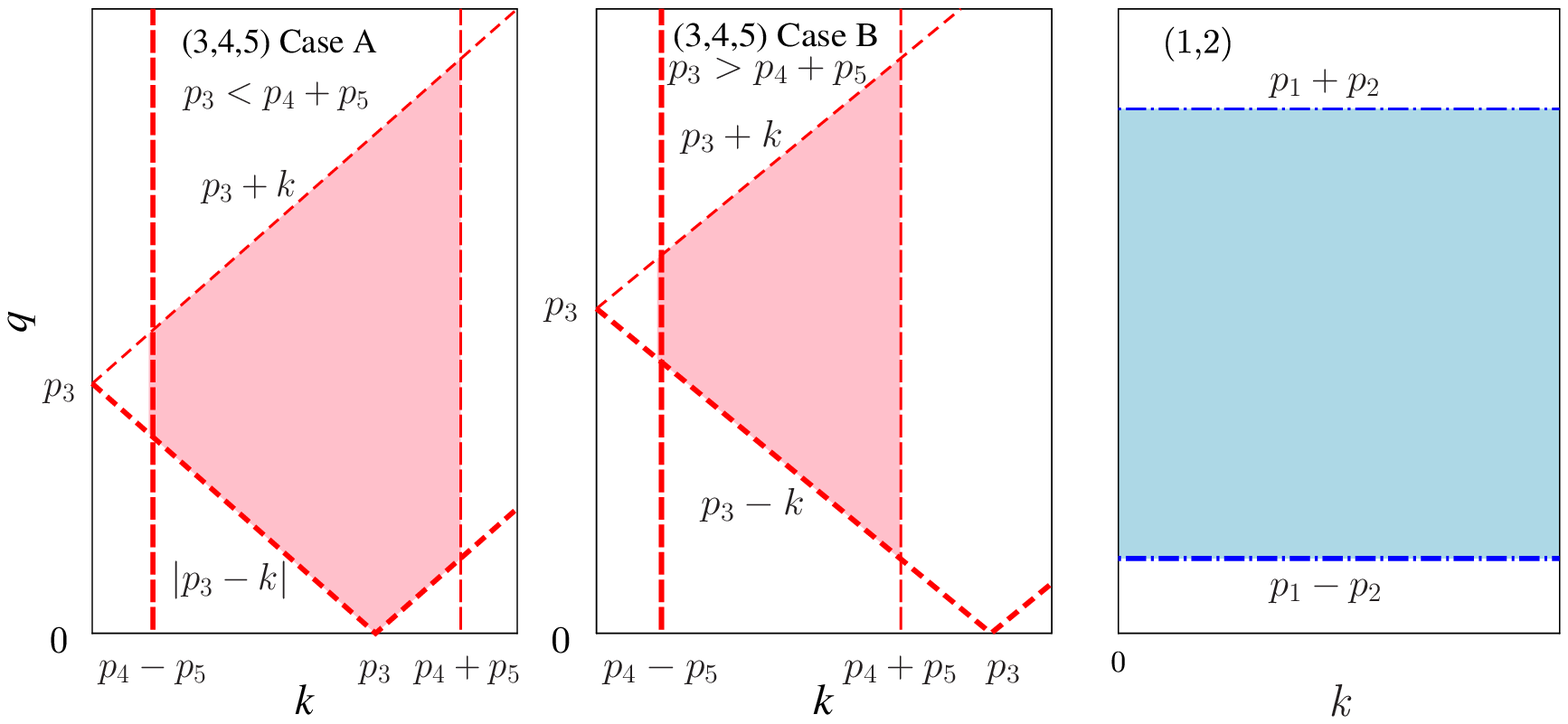}%{graphaz.eps}%
\caption{\label{f:regions}
Schematic representation of dashed areas in the $q-k$ plane
restricted by momentum transfers in subsystem (3,4,5) at
$p_3<p_4+p_5$ (left, case A) and at $p_3>p_4+p_5$ (middle, case B),
as well as in subsystem (1,2) (right). $Q^{(5)}$ is the intersection
area $\Sigma$ for the subsystems (3,4,5) and (1,2). In this and
other figures thin short-dashed lines refer to $q=p_3+k$; thick
short-dashed lines to $q=|p_3-k|$; thick long-dashed lines to
$k=p_4-p_5$, thin long-dashed lines to $k=p_4+p_5$; thick dot-dashed
line to $q=p_1-p_2$, while thin dot-dashed line to $q=p_1+p_2$.}
\end{figure*}

Then, following Eq.~(\ref{e:n=4}), one can introduce  the second
auxiliary momentum $\bm{k}$. It occurs due to an additional degree
of freedom associated with the third particle in the subsystem
(3,4,5). Using again Eq.~(\ref{e:n=3}) we obtain
\begin{eqnarray}
%\hspace{-2.5cm}
& &  I_\Omega^{(5)}(p_1,p_2,p_3,p_4,p_5) \nonumber     \\
& &  {\dg =}{1 \over (4\pi)^2}\ \int q^2 {\rm d}q \int k^2 {\rm d}k\
           I_\Omega^{(3)}(p_1, p_2; q)\,
           I_\Omega^{(3)}(p_3; q, k)\,
\nonumber \\
& &    \times       I_\Omega^{(3)}(p_4, p_5; k)
= \frac{2 (2\pi)^4}{p_1 p_2 p_3 p_4 p_5}\,  Q^{(5)},
\label{e:n=5}
\end{eqnarray}
where
\begin{equation}
  Q^{(5)}=  \int_{\Sigma} {\rm d}q\,{\rm d}k=\Sigma,
\label{e:Q5}
\end{equation}
with $\bm{k}=\bm{p}_4+\bm{p}_5$ and $\bm{q}=\bm{p}_3+\bm{k}$.
Therefore, $Q^{(5)}$ is equal to the area { $\Sigma$} in the
$(q,k)$-plane (Fig.\ \ref{f:regions}) restricted by the conditions
\begin{eqnarray}
p_1-p_2 < q < p_1 + p_2 & {\rm ~~~ for~fermions~(1,2);}
 \label{e:12}
\end{eqnarray}
\begin{eqnarray}
 |k-p_3| < q < p_3+k,& p_4-p_5 < k < p_4+p_5
\label{e:345}
\end{eqnarray}
for~fermions~(3,4,5).

\begin{figure*}[htb!]
\centering
\includegraphics[width=0.90\textwidth]{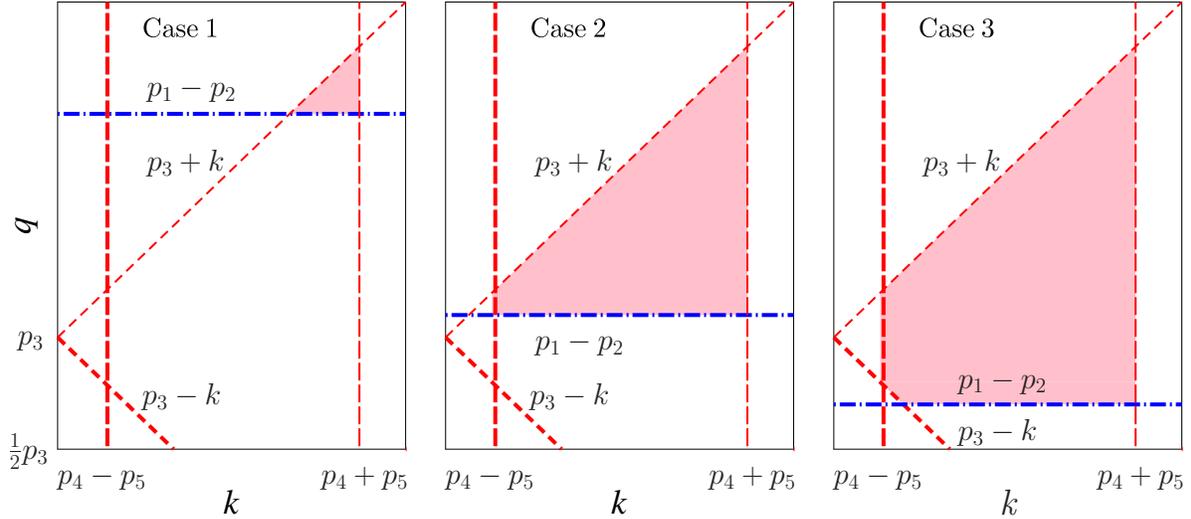}%{grazzz.eps}%
\caption{\label{f:regions123}
Schematic representation of (shaded) areas $Q^{(5)}=\Sigma$ for
cases 1, 2 and 3.}
\end{figure*}

\begin{figure*}[htb!]
\centering
\includegraphics[width=0.97\textwidth, bb= 0 0 550 230]{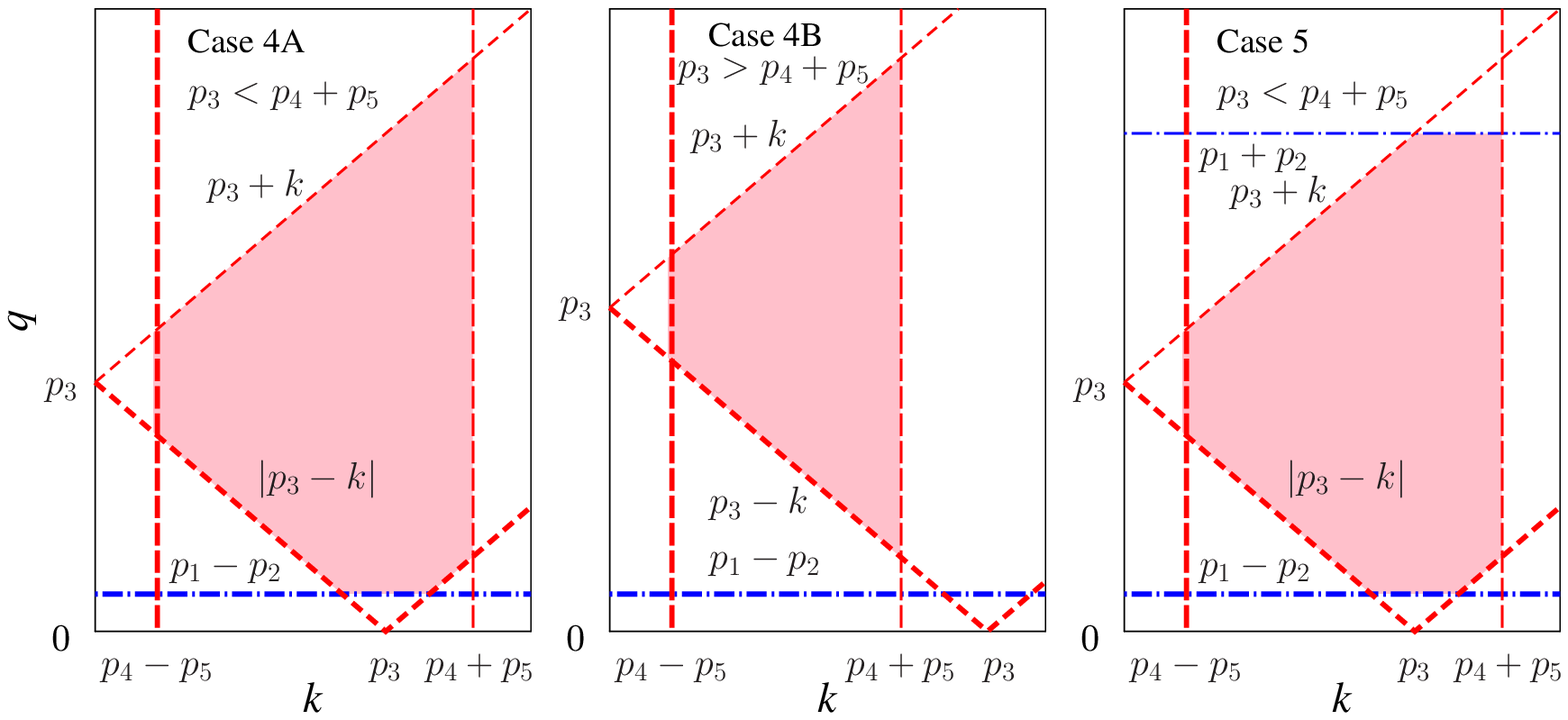}%{graphazz4aa.eps}%{graphaz4ab.eps}%
\caption{\label{f:regions4}
Schematic representation of areas $Q^{(5)}=\Sigma$ for cases 4A, 4B
and 5. }
\end{figure*}

The required area $Q^{(5)}$   in the $(q,k)$  plane can be
calculated from a simple geometrical consideration (Fig.\
\ref{f:regions}). Generally, one can distinguish six different cases
which we denote as cases 1, 2, 3, 4A, 4B and 5. Geometry for these
cases is presented in Figs.\ \ref{f:regions123} and
\ref{f:regions4}. The conditions for the realization of these cases
and relevant expressions for $Q^{(5)}$ are listed in
Table~\ref{tabQ5}. In Fig.\ \ref{f:regions123} we do not show the
region of low $q$ because the area  $Q^{(5)}$ lies above that
region. In all the cases 1 -- 4 the line $q=p_1+p_2$ is placed too
high and does not affect { the value of } $Q^{(5)}$; accordingly, we
do not plot this line in such cases.

We have enumerated six cases in Table \ref{tabQ5}, in descending
order in  $p_1$,  which is seen from column 2. Replacing the
inequalities written in this column by the equalities we obtain the
boundaries of corresponding regions in the allowed parameter space
of $p_1,\ldots p_5$. The six regions fully cover the allowed
parameter space, and $Q^{(5)}$ changes continuously while going from
one region to another. The two subcases, 4A and 4B, correspond to
the conditions $p_3<p_4+p_5$ and $p_3>p_4+p_5$, respectively, where
the area $\Sigma$ is calculated differently (see the left and middle
panels in Fig.\ \ref{f:regions4}). The conditions $p_3<p_4+p_5$ and
$p_3>p_4+p_5$ are not written explicitly in column 2 for  { the
cases}  A and B because they are guaranteed by the inequalities
written for these cases in column 2  $(p_1 \geq p_2)$. In  { the}
case 5 one would also expect two similar subcases, A and B, but the
subcase B, in which one would have $p_1<-p_2+p_3+p_4+p_5$, cannot be
realized because it is incompatible with $p_3>p_4+p_5$\ ($p_1+p_2
\geq 2p_3$). Therefore, case 5 is essentially the same as subcase
5A.

%%%%%%%%%%%%%%%%%%%%%%%%%%%%%%
%%%%%%%%%%%%%%%%%%%%%%%%%%%%%%%%

\begin{table*}
%T2%
\small \caption{Values of $Q^{(5)}$ for all six cases displayed in
Figs.\ \ref{f:regions123} -- \ref{f:regions4}} \label{tabQ5}
\begin{tabular}{clll}
\hline \hline
Case & Conditions & & $Q^{(5)}$ \\
\hline
1  & $p_1<p_2+p_3+p_4+p_5;$ & $p_1>p_2+p_3+p_4-p_5$ &
 $\frac{1}{2}(p_2+p_3+p_4+p_5-p_1)^2$ \\
2 & $p_1<p_2+p_3+p_4-p_5;$ &$p_1>p_2+p_3-p_4+p_5$ & $2p_5(p_2+p_3+p_4-p_1)$ \\
3 & $p_1<p_2+p_3-p_4+p_5;$ & $p_1> p_2+|p_3-p_4-p_5|$&
$4p_4p_5-{1 \over 2}(p_1+p_4+p_5-p_2-p_3)^2$ \\
4A & $p_1<p_2-p_3+p_4+p_5;$ & $p_1>-p_2+p_3+p_4+p_5$ & $4p_4p_5-(p_4+p_5-p_3)^2-(p_1-p_2)^2$ \\
4B & $p_1<p_2+p_3-p_4-p_5;$ & $p_1>-p_2+p_3+p_4+p_5$ & $4p_4p_5$ \\
5 &$p_1<-p_2+p_3+p_4+p_5;$ & $p_1 \geq p_2$ &
$4p_4p_5-(p_4+p_5-p_3)^2-(p_1-p_2)^2$ \\
  &  &   &    $- \frac{1}{2} (p_3+p_4+p_5-p_1-p_2)^2$    \\
\hline \hline
\end{tabular}
%\item[] $^{\rm a}$ Self-supporting.
%\item[] $^{\rm b}$ Deposited over Al backing.
%\end{indented}
\end{table*}
%%%%%%%%%%%%%%%%%%%%%%%%%%%%%%%%%%

%%%%%%%%%%%%%%%%%%%%%%%%%%%%%%%%%%
%%%%%%%%%%%%%%%%%%%%%%%%%%%%%%%%%%%

%%%%%%%%%%%%%%%%%%%%%%%%%%%%%%%%%%%
%%%%%%%%%%%%%%%%%%%%%%%%%%%%%%%%%%%%

Table~\ref{tabQ5} gives $Q^{(5)}$ for any system of five degenerate
fermions. Furthermore, the results are simplified if several
fermions have equal momenta. For the convenience of the reader we
present such results in Tables~\ref{tab3} and \ref{tab4}. Notice
that, as a rule, the number of possible cases for equal momenta of
fermions becomes lower (not all the cases are necessarily realized).
In the simplest case of five equal momenta $p_1=p_2=p_3=p_4=p_5$,
one has (case 5) $Q^{(5)}=\frac{5}{2}\,p_1^2$.

%TA1%
\begin{table*}
\small \caption{Values of $Q^{(5)}$ for two equal fermions}
\label{tab3}
\begin{tabular}{cll}
\hline \hline
Case &   Conditions   & $Q^{(5)}$\\
\hline
4A & $p_1=p_2$,~~$p_3<p_4+p_5$,~$2p_1>p_3+p_4+p_5$ & $4p_4p_5-(p_4+p_5-p_3)^2$ \\
%\mr
4B &  $p_1=p_2$,~~$p_3>p_4+p_5$,~$2p_1>p_3+p_4+p_5$ & $4p_4p_5$ \\
%\mr
5 & $p_1=p_2$,~~$2p_1<p_3+p_4+p_5$  & $4p_4p_5-(p_4+p_5-p_3)^2$ \\
 && $-{1 \over 2}(p_3+p_4+p_5-2p_1)^2$  \\
\hline
1 & $p_2=p_3$,~~$p_1< 2p_2+p_4+p_5 $,~$p_1>2p_2+p_4-p_5$
&$\frac{1}{2}(2p_2+p_4+p_5-p_1)^2$ \\
%\mr
2 &  $p_2=p_3$,~~$p_1<2p_2+p_4-p_5$,~$p_1>2p_2-p_4+p_5$ &$2p_5(2p_2+p_4-p_1)$ \\
%\mr
3 & $p_2=p_3$,~~$p_1<2p_2-p_4+p_5$,~$p_1>p_2+|p_2-p_4-p_5|$&
$4p_4p_5-{1 \over 2}(p_1+p_4+p_5-2p_2)^2$ \\
%\mr
4B &$p_2=p_3$,~~$p_1<2p_2-p_4-p_5$,~$p_1>p_4+p_5$ & $4p_4p_5$ \\
%\mr
5 & $p_2=p_3$,~~$p_1<p_4+p_5$  & $4p_4p_5-(p_4+p_5-p_2)^2-(p_1-p_2)^2$ \\
    &  &  $-{1 \over 2}(p_4+p_5-p_1)^2$ \\
\hline
1 & $p_3=p_4$,~~$p_1<p_2+2p_3+p_5$,~ $p_1>p_2+2p_3-p_5$ & $\frac{1}{2}(p_2+2p_3+p_5-p_1)^2$ \\
%\mr
2 & $p_3=p_4$,~~$p_1<p_2+2p_3-p_5$,~$p_1>p_2+p_5$ & $2p_5(p_2+2p_3-p_1)$ \\
%\mr
4A & $p_3=p_4$,~~$p_1< p_2+p_5$,~$p_1>-p_2+2p_3+p_5$ & $4p_3p_5-p_5^2-(p_1-p_2)^2$ \\
%\mr
5 & $p_3=p_4$,~~$p_1< -p_2+2p_3+p_5$  & $4p_3p_5-p_5^2-(p_1-p_2)^2-{1 \over 2}(2p_3+p_5-p_1-p_2)^2$ \\
\hline
1 &  $p_4=p_5$,~~$p_1<p_2+p_3+2p_4$,~$p_1>p_2+p_3$ & $\frac{1}{2}(p_2+p_3+2p_4-p_1)^2$ \\
%\mr
3 &  $p_4=p_5$,~~$p_1<p_2+p_3 $,~ $p_1>p_2+|p_3-2p_4|$ & $4p_4^2-{1 \over 2}(p_1+2p_4-p_2-p_3)^2$ \\
%\mr
4A &  $p_4=p_5$,~~$p_1<p_2-p_3+2p_4 $,~ $p_1>-p_2+p_3+2p_4$ & $4p_4^2-(2p_4-p_3)^2-(p_1-p_2)^2$ \\
%\mr
4B &  $p_4=p_5$,~~$p_1<p_2+p_3-2p_4$,~$p_1>-p_2+p_3+2p_4$ & $4p_4^2$  \\
%\mr
5 &  $p_4=p_5$,~~$p_1<-p_2+p_3+2p_4$              & $4p_4^2-(2p_4-p_3)^2-(p_1-p_2)^2$ \\
&&$-{1\over 2}(p_3+2p_4-p_1-p_2)^2$\\
\hline \hline
\end{tabular}
%\item[] $^{\rm a}$ Self-supporting.
%\item[] $^{\rm b}$ Deposited over Al backing.
%\end{indented}
\end{table*}
%%%%%%%%%%%%%%%%%%%%%%%%%%%%%%%%%%%%%%%%%%%%%%%%%%%%%%%%%%%%

%%%%%%%%%%%%%%%%%%%%%%%%%%%%%%%%%%%%%%%%%%%%%%%%%%%%%%%%%%%%%
%TA5%
\begin{table*}
\small \caption{Values of $Q^{(5)}$ for more than two particles with equal
momenta} \label{tab4}
\begin{tabular}{clll}
\hline \hline
Case &     Conditions &  & $Q^{(5)}$\\
\hline
4A & $p_1=p_2,$ $p_3=p_4$    & $2p_1 > 2p_3+p_5$ & $4p_3p_5-p_5^2$ \\
%\mr
5 &  $p_1=p_2,$ $p_3=p_4$   & $2p_1< 2p_3+p_5$ &
$4p_3p_5-p_5^2 -{1 \over 2}(2p_3+p_5-2p_1)^2$ \\
\hline
4A   & $p_1=p_2,$ $p_4=p_5$ &$p_3<2p_4$,~ $2p_1> p_3+2 p_4$ & ${ 4p_4^2}-(2p_4-p_3)^2$ \\
%\mr
4B & $p_1=p_2,$ $p_4=p_5$  &$p_3 > 2p_4$,~  $2p_1> p_3+2p_4$ &  $4p_4^2$  \\
%\mr
5 & $p_1=p_2,$  $p_4=p_5$  & $2p_1< p_3+2p_4$ & $4p_4^2-(2p_4-p_3)^2- {1 \over 2}(p_3+2p_4-2p_1)^2$ \\
\hline
1 & $p_2=p_3,$ $p_4=p_5$ & $p_1<  2p_2+2p_4$,~ $p_1>2p_2$ & $\frac{1}{2}(2p_2+2p_4-p_1)^2$ \\
%\mr
3 & $p_2=p_3,$ $p_4=p_5$ & $p_1< 2p_2$,~ $p_1>p_2+|p_2-2p_4|$  & $4p_4^2-{1 \over 2}(p_1+2p_4-2p_2)^2$ \\
%\mr
4B & $p_2=p_3,$ $p_4=p_5$ & $p_1<2p_2-2p_4$,~ $p_1>2p_4$ &  $4p_4^2$  \\
%\mr
5 & $p_2=p_3,$ $p_4=p_5$ & $p_2< p_1 <2p_4$ & $4p_4^2-(2p_4-p_2)^2-{1 \over 2}(2p_4-p_1)^2-(p_1-p_2)^2$ \\
\hline
4B & $p_1=p_2=p_3$& $p_1> p_4+p_5$ & $4p_4p_5$ \\
%\mr
5 &  $p_1=p_2=p_3$& $p_1< p_4+p_5$ & $4p_4p_5-{3 \over 2}(p_4+p_5-p_1)^2$ \\
\hline
4B & $p_1=p_2=p_3,~p_4=p_5$  & $p_1> 2p_4$ & $4p_4^2$ \\
%\mr
5 & $p_1=p_2=p_3,~p_4=p_5$   & $p_1< 2p_4$ & $4p_4^2-{3 \over 2}(2p_4-p_1)^2$ \\
\hline
1   & $p_2=p_3=p_4$ & $p_1< 3p_2+p_5$,~ $p_1>3p_2-p_5$ & ${1 \over 2}(3p_2+p_5-p_1)^2$ \\
%\mr
2 & $p_2=p_3=p_4$ &   $p_1<3p_2-p_5$,~ $p_1>p_2+p_5$ &  $2p_5(3p_2-p_1)$  \\
%\mr
5 & $p_2=p_3=p_4$ & $p_1 < p_2+p_5$ & $p_1p_5+{3 \over 2}p_1(2p_2-p_1)-{3 \over 2}(p_2-p_5)^2$ \\
\hline
1   & $p_3=p_4=p_5$ & $p_1 < p_2+3p_3$,~$p_1>p_2+p_3$ & ${1 \over
2}(p_2+3p_3-p_1)^2$ \\
%\mr
4A & $p_3=p_4=p_5$ &  $p_1 < p_2+p_3 $,~$p_1>-p_2+3p_3$ &  $3p_3^2-(p_1-p_2)^2$  \\
%\mr
5 & $p_3=p_4=p_5$ & $p_1<-p_2+3p_3$   &
$3p_3^2-(p_1-p_2)^2 - {1 \over 2}(3p_3-p_2-p_1)^2$ \\
\hline
4A &  $p_1=p_2>p_3=p_4=p_5$ & $2p_1>3p_3$ &  $3p_3^2$  \\
%\mr
5 & $p_1=p_2>p_3=p_4=p_5$ & $2p_1<3p_3$   &
$3p_3^2 - {1 \over 2}(3p_3-2p_1)^2$ \\
\hline
1    &$p_2=p_3=p_4=p_5$ &  $2p_2< p_1 < 4p_2$ & ${1 \over 2}(4p_2-p_1)^2$ \\
%\mr
5   & $p_2=p_3=p_4=p_5$ &   $p_1 < 2p_2$ &  $4p_1p_2-{3 \over 2}p_1^2$  \\
%\br
5   &  $p_1=p_2=p_3=p_4$ & $p_1>p_5$  & $4p_1p_5-{3 \over 2}p_5^2$ \\
\hline
%\br
5   & $p_1=p_2=p_3=p_4=p_5$ & & ${5 \over 2} p_1^2$ \\
\hline \hline
\end{tabular}
%\item[] $^{\rm a}$ Self-supporting.
%\item[] $^{\rm b}$ Deposited over Al backing.
%\end{indented}
\end{table*}
%%%%%%%%%%%%%%%%%%%%%%%%%%%%%%%%%%%%%%%%%%%%%%%%%%%%%%%%%%%%%

\begin{figure}[htb!]
\centering
\includegraphics[width=0.45\textwidth]{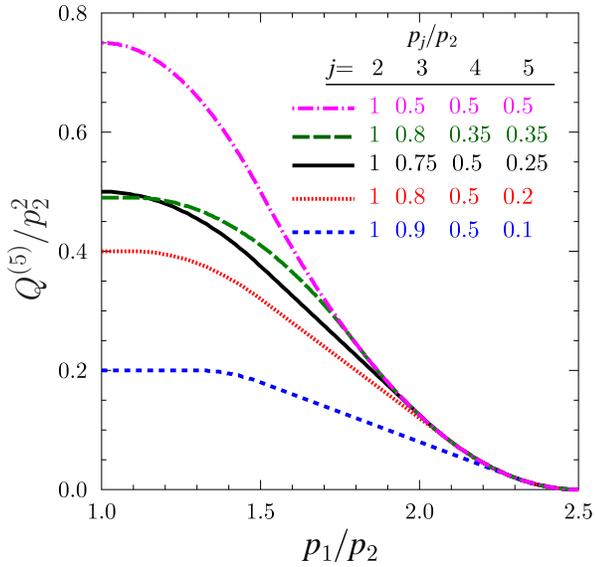}%{q5aaa.eps}%
\caption{\label{f:q5}
Dependence of $Q^{(5)}/p_2^2$ on $p_1/p_2$ (five lines of different
types) for five different combinations of $p_2,\ldots,p_5$ indicated
in the inserted Table;
%{\dg  $p_{1 \rm max}=2.5 p_2$ for all combinations}
$p_{1 \rm max}=2.5 p_2$ for all combinations
(see text
for details).}
\end{figure}

For illustration, Fig.\ \ref{f:q5} presents $Q^{(5)}$ as a function
of $p_1$ for five combinations of $p_j$ ($j=$2,\dots,5).  To exhibit
dimensionless quantities we plot $Q^{(5)}/p_2^2$ versus $p_1/p_2$
(five lines of different types). Any line refers to a certain
combination of $p_j/p_2$ listed in the table that is inserted in the
figure. For the convenience of presentation, the values of $p_j$ are
chosen in such a way that $p_{1 \rm max}=p_2+p_3+p_4+p_5=2.5\,p_2$
for each combination, so that $p_1$ varies from $p_2$ to $2.5p_2$.
The solid, dotted and short-dashed lines correspond to the cases in
which all $p_2,\ldots,p_4$ are different. The long-dashed line is
for a pair of equal momenta ($p_4=p_5=0.35p_2$), while the
dot-dashed line is for three equal momenta ($p_3=p_4=p_5=0.5p_2$).
Naturally, $Q^{(5)}\to 0$ as $p_1 \to p_{1 \rm max}$ for all lines
{(case 1)}. When $p_1$ decreases from $p_{1 \rm max}$, the quantity
$Q^{(5)}$ grows up owing to the increase of possible configurations
of particle momenta allowed by momentum conservation.

%%%%%%%%%%%%%%%%%%%%%%%%%%%%%%%%%%%%%
%%%%%%%%%%%%%%%%%%%%%%%%%%%%%%%%%%%%%

Using similar technique one can calculate angular integrals
$Q^{(n)}_\Omega$ at higher $n>5$. With increasing $n$ the number of
different cases will be progressively larger.

%%%%%%%%%%%%%%%%%%%%%%%%%%%%%%%%%%%%%%%%%%%
\section{Discussion}
\label{s:applic}

Let us outline some applications of our results. A very rich
spectrum of applications is provided by neutrino physics of neutron
star cores (Section \ref{s:introduc}) where many neutrino mechanisms
can operate and regulate thermal evolution of these stars (e.g.,
\citealt{YKGH01}). The neutrino emissivities of many of these reactions,
especially in nucleon-hyperon matter, have not been calculated
with sufficient accuracy. The detailed calculations require the above
results.

The major neutrino processes in neutron star cores can be divided
into (i) direct Urca processes, (ii) baryon-baryon bremsstrahlung
processes, and (iii) modified Urca processes. The approximation of
angle-independent squared matrix elements (Section \ref{s:introduc})
for these reactions is usually valid \citep{YKGH01} which justifies
our analysis. Direct Urca processes are much stronger, than other
ones, but have a threshold character (e.g. \citealt{LPPH91,PPLP92}).
Any direct Urca process switches on once the density exceeds some
threshold value, $\rho=\rho_{\rm th}$, determined by a given process
and a given equation of state of the matter. As a rule, threshold
densities $\rho_{\rm th}$ lie in the inner cores of massive neutron
stars. If the direct Urca processes are allowed, the modified Urca
and the bremsstrahlung processes are insignificant. Some equations
of state forbid the onset of direct Urca processes at any density of
matter in stable neutron stars.

Following \citet{YKGH01}, we illustrate the variety of
neutrino emission mechanisms in a neutron star core using a model of
nucleon-hyperon matter as an example and assuming the presence of
$\Lambda$ and $\Sigma^-$ hyperons. Generally, the Fermi momenta of
constituents of the matter (n, p, e, $\mu$, $\Lambda$, $\Sigma$) may
be rather arbitrary (e.g., see  \citealt{HPY07} and references
therein). For simplicity, we consider the baryons (n, p, $\Lambda$,
$\Sigma$) as non-superfluid and assume the presence of beta
equilibrium and electric neutrality \citep{ST83,HPY07}.

\subsection{Direct Urca processes ($n$=3)}
\label{s:DU}

Any direct Urca process is a sequence of two reactions,
\begin{equation}
  B_a \to B_b+\ell+\overline{\nu}_\ell, \quad B_b+ \ell \to B_a+\nu_\ell,
\label{e:DU}
\end{equation}
which results in the emission a neutrino pair. Here, $\ell$ stands
for a lepton (e or $\mu$); $\nu_\ell$ and $\overline{\nu}_\ell$ are
associated neutrino and anti-neutrino, respectively; $B_a$ and $B_b$
are baryons which undergo lepton decay or capture. In our example,
we have four baryon pairs (e.g. \citealt{PPLP92}) ($B_a,~B_b$)=(n,
p), ($\Lambda$, p), ($\Sigma$, n) and ($\Sigma,~\Lambda$), and,
hence, eight direct Urca processes ($\ell=$ e or $\mu$).  The
angular integrals for both reactions in (\ref{e:DU}) are evidently
equal. They  involve three fermions (Section \ref{s:n*nd3}) and are
given by Eq.\ (\ref{e:n=3}). Any direct Urca process operates as
long as $p_1 \leq p_2+p_3$, and the equality $p_1=p_2+p_3$
determines its threshold density $\rho_{\rm th}$. The emissivities
of the direct Urca processes in the nucleon-hyperon matter have been
calculated by \citet{LPPH91,PPLP92}.

\subsection{Bremsstrahlung processes ($n$=4)}
\label{s:Brems}

These are the processes of nucleon-nucleon collisions accompanied by
the emission of neutrino-pairs (any flavors, $\nu_{\rm e}$,
$\nu_{\mu}$, $\nu_\tau$). The angular integrals involve four
fermions (Section \ref{s:n=4}). Such processes can be divided into
three types (e.g. \citealt{M87}).

The bremsstrahlung processes of the first type are
\begin{equation}
   B+B \to B+B+\nu+\overline{\nu},
\label{e:Br1}
\end{equation}
where $B$=n, p, $\Sigma$ or $\Lambda$ is any baryon (with Fermi
momentum $p_1$). We have four such processes. Accor\-ding to the
results of Section \ref{s:n=4}, in this case
\begin{equation}
    I_{\Omega}^{(4)}= {4\ (2\pi)^3 \over p_1^3}.
\label{e:IBr1}
\end{equation}

The processes of the second type read
\begin{equation}
   B_1+B_2 \to B_1+B_2+\nu+\overline{\nu},
\label{e:Br2}
\end{equation}
with $p_1>p_2$. We have six such processes for which
\begin{equation}
    I_{\Omega}^{(4)}= {4\ (2\pi)^3 \over p_1^2 p_2}.
\label{e:IBr2}
\end{equation}
Eqs. (\ref{e:IBr1}) and (\ref{e:IBr2}) are in line with the formulas
presented in the literature (e.g., \citealt{YKGH01}) for the
nn, pp and np bremsstrahlung processes.

Finally, we have two other bremsstrahlung processes with four
different fermions,
\begin{equation}
  \Lambda+{\rm n} \to \Sigma + {\rm p} + \nu+\overline{\nu},\quad
  \Sigma+{\rm p} \to \Lambda + {\rm n} + \nu+\overline{\nu}.
\end{equation}
Their angular integrals are easily calculated from Eqs.
(\ref{e:n=4}) and (\ref{e:4diff}). They are given by different
expressions depending on the values of four Fermi momenta.

The neutrino emissivities $Q$ of the bremmstrahlung processes in
nucleon matter have been studied with considerable attention (e.g.
\citealt{YKGH01} and references therein), but the emissivities of
the processes involving hyperons are much less elaborated. Their
calculation would require the angular integrals $Q^{(4)}$ of Section
\ref{s:n=4} (Table \ref{tabQ4}).

\subsection{Modified Urca processes ($n$=5)}
\label{s:MU}

Modified Urca processes are similar to direct Urca ones, Eq.\
(\ref{e:DU}) (Section \ref{s:DU}), but involve an additional baryon
$B_c$ in the initial and final channels,
\begin{equation}
  B_a+B_c \to B_b+B_c+\ell+\overline{\nu}_\ell,
    \quad B_b+B_c + \ell \to B_a+B_c+\nu_\ell.
\label{e:MU}
\end{equation}
Here, $B_a$ and $B_b$ are the same as in Eq.\ (\ref{e:DU}), while
$B_c$ can be any baryon available in the matter. In our example we
have $8 \times 4=32$ modified Urca processes and we need a
five-fermion angular integral $I_\Omega^{(5)}$ for each of them. As
in Section \ref{s:DU}, the angular integrals for both reactions in
(\ref{e:MU}) are equal. The required integrals are presented in
Section \ref{s:n=5},  Tables \ref{tabQ5}--\ref{tab4}. We can
generally divide all these modified Urca processes into two groups.

In the first group we include all the processes with $B_c$ equal to
either $B_a$ or $B_b$. These are five-fermion processes with three
identical fermions. Their angular integrals are given by Eq.\
(\ref{e:n=5}) with $Q^{(5)}$ listed in Table \ref{tab4}, depending
on Fermi momenta of reacting particles.

The second group contains other modified Urca processes with only
two  identical  fermions $B_c$ out of five. Corresponding angular
integrals are also described by Eq.\ (\ref{e:n=5}) with $Q^{(5)}$
presented in Table \ref{tab3}.

Consider, for instance, the simplest case of nucleon dense matter
which consists of n, p, e and $\mu$. In realistic models of such a
matter in neutron stars cores the neutrons dominate (the neutron
number density $n_{\rm n}$ is the largest one), and the electric
neutrality of the matter implies that the number densities of other
particles satisfy the condition $n_{\rm p}=n_{\rm e}+n_{\mu}$. Then
the Fermi momenta of the particles obey the inequality $p_{\rm
n}>p_{\rm p}>p_\ell$. In this case we have four modified Urca
processes. They are the neutron-branch
(n+n$\to$p+n+$\ell+\overline{\nu}_\ell$, p+n+$\ell \to$n+{
n}+${\nu}_\ell$, $B_c$=n) and proton-branch
(n+p$\to$p+p+$\ell+\overline{\nu}_\ell$, p+p+$\ell
\to$n+p+${\nu}_\ell$, $B_c$=p) processes with electrons or muons
($\ell$=e or $\mu$).

Consider the neutron branch of the process using Table~\ref{tab4}
($p_1=p_2=p_3$).
In {the} case 4B one has $p_{\rm n}> p_{\rm p}+p_\ell$ meaning that
the direct Urca process is forbidden (Section \ref{s:DU}). Then we
obtain
\begin{equation}
    I_{\rm \Omega}^{(5)}={8 (2\pi)^4 \over p_{\rm n}^3},
\label{T10_4B}
\end{equation}
which is in line with Eq.~(F.11) of \citet{ST83} or Eq.~(134)
of \citet{YKGH01}.

In the case 5 we have $p_{\rm n} <  p_{\rm p} +  p_{\ell}$, i.e.
the direct Urca process is allowed. Then we have
\begin{equation}
    I_\Omega^{(5)}=\frac{8 (2\pi)^4}{p_{\rm n}^3} \,
     \left[ 1- {3 \over 8} \, \frac{ ( p_{\rm p} +  p_\ell - p_{\rm n})^2 }
     {p_{\rm p} p_\ell }  \right].
\label{T10_5A}
\end{equation}
However, this angular integral is rather unimportant because the modified
Urca process is insignificant when the direct Urca operates.

For the proton branch of the modified Urca process we also use
Table~\ref{tab4} ($p_2=p_3=p_4$). In { the} cases 1 and 2 the condition $p_{\rm n}
> p_{\rm p} +  p_{\ell}$ is satisfied and the direct Urca process is
forbidden. In the case 1 at $p_{\rm n} > 3p_{\rm p} - p_{\ell}$ we
have
\begin{equation}
 I_\Omega^{(5)}={(2\pi)^4 \over p_{\rm n} p_{\rm p}^3 p_\ell} \,
                     (3p_{\rm p}+p_\ell-p_{\rm n})^2,
\label{T11_1}
\end{equation}
which agrees with Eq.~(141) of \citet{YKGH01} but slightly
differs from Eq.~(29) of \citet{HLY01}, where one factor
$p_{\rm p}$ in the denominator is erroneously replaced by  $p_{\rm
n}$.

In the case 2 at $3p_{\rm p} - p_\ell > p_{\rm n} > p_{\rm
p}+p_{\ell}$ we obtain, in agreement with  Eq.~(35) in
\citet{HLY02},
\begin{equation}
 I_\Omega^{(5)}={4(2\pi)^4 \over p_{\rm n} p_{\rm p}^3}\,
                     (3p_{\rm p}-p_{\rm n}).
\label{T11_2}
\end{equation}

If the direct Urca process is open, $p_{\rm n} < p_{\rm
p}+p_{\ell}$, we come to {the} case  5 in which
%%%%%%%%%%%%%%%%%%%%%%%%%
\begin{equation}
  I_\Omega^{(5)}={2 (2\pi)^4 \over p_{\rm p}^3} \,
     \left[ 1 + {3 \over 2}\ { 2 p_{\rm p} - p_{\rm n}
     \over  p_{\ell} }  - {3 \over 2}
          {(p_{\rm p}-p_{\ell})^2 \over p_{\rm n}p_{\ell}}
           \right].
\label{T11_5A}
\end{equation}
This case, omitted in literature, is insignificant because the
direct Urca process is much stronger than the modified one.

As for the numerous modified Urca processes with hyperons, their
study has to be performed accurately based on the results of Section
\ref{s:n=5}.

\subsection{Other applications}

In solid-state physics (metals, degenerate semi\-con\-ductors) one
often needs \citep{Ziman,Kittel} effective collision frequencies of
strongly degenerate electrons (e+e$\to$e+e) which contain
$I_\Omega^{(4)}$ (Section \ref{s:n=4}) with four equal Fermi momenta
of strongly degenerate electrons, $p_{\rm Fe}\equiv p_{\rm e}$.
These collision frequencies determine kinetic coefficients due to
electron-electron collisions. Using Eq.~(\ref{e:n=4}) and the last
line in Table~\ref{tabQ4} we immediately obtain the well known
result $I_\Omega^{(4)}= 4\ (2\pi)^3 / p_\mathrm{e}^3$ (e.g.,
\citealt{Ziman}). The variety of similar applications in different
studies of strongly degenerate fermionic systems is very large.

%%%%%%%%%%%%%%%%%%%%%%%%%%%%%%%%%%%%%%%%%%%%%%
\section{Conclusions}
\label{s:conclusion}

We have described calculations of angular integrals
$I^{(n)}_\Omega$, Eq.\ (\ref{e:I1}), which determine neutrino emissivities,
reaction rates and related quantities for reactions involving $n$ degenerate
fermions (in initial + final channels)  with  different or
equal Fermi momenta $p_1,\ldots,p_n$. These angular integrals often
occur in applications if differential reaction probabilities are
determined by angle-averaged squared matrix elements (Section
\ref{s:introduc}). The advantage of angular integrals
$I^{(n)}_\Omega$ is that they solely depend on $n$ Fermi-momenta
$p_1,\ldots,p_n$, being independent of the nature of reacting
fermions and their interactions. The integrals $I^{(n)}_\Omega$ are
described by analytic expressions which may have different forms,
but they can be calculated once and forever.

We have calculated $I^{(n)}_\Omega$ for all possible cases with
$n$=2 and 3 (Section \ref{s:n*nd3}), 4 (Section \ref{s:n=4}) and 5
(Section \ref{s:n=5}). The formalism we have used (Section
\ref{s:remarks}) allows one to perform similar calculations for
higher~$n$.

In Section \ref{s:applic} we have outlined some applications of the
results, particularly, for neutrino emission processes in neutron
star cores composed of nucleons and hyperons. For illustration, we
have discussed the expressions for angular integrals of major
neutrino emission processes in neutron star cores containing
neutrons, protons, electrons, muons, as well sigma and lambda
hyperons. They are eight direct Urca processes ($n=3$), 12
baryon-baryon bremsstrahlung processes ($n=4$) and 32 modified Urca
processes ($n=5$). The majority of these neutrino reactions have not
been studied with considerable attention. We provide the angular
integrals which are the most important ingredients for such studies.
Our results can be useful for construc\-ting a uniform database of
neutrino emissivities in nucle\-on-hyperon matter of neutron star
cores which is needed to simulate thermal structure and evolution of
neutron stars.

Let us stress that much work is required to complete such a
database. Aside of the angular integrals calculated here, one needs
the matrix elements of many neutrino reactions as well as the
factors which describe the suppression of these reactions by
possible superfluidity of nucleons and hyperons.
This suppressi\-on
can be either very strong or weak depending on (largely unkown)
critical temperatures for superfluidity of different particles
%%(as we know from studies of superfluid nucleon pairs {cores?},
(e.g., \citealt{YKGH01}). It would be a complicated project to
calculate the matrix elements and suppression factors from first
principles but we expect to simplify this task using some
selfsimilarity criteria, like those formulated, in \citet{YKGH01}.
In addition, superfluidity of various baryon species can induce a
specific neutrino emission due to Cooper pairing of baryons. Such
processes involving hyperons should also be studied and included
into the database
%%which is a separate complicated problem because of
%%delicate collective
taking into account in-medium effects in systems of superfluid
baryons (\citealt{lp06}; also see references given by
\citealt{pageetal2011,shterninetal2011}).

Note  also, that neutrino reactions can be affected by strong
magnetic fields. Much work should be done to study the effects of
magnetic fields on various neutrino processes. The available
calculations of these processes in magnetized neutron star crust and
nucleon core (reviewed by \citealt{YKGH01}) show that one typically
needs very strong fields to affect the neutrino emission of neutron
stars. For instance, as demonstrated by \citet{BY99}, the direct
Urca process in nucleon neutron star core can be noticeably affected
by the fields $B \gtrsim 10^{16}$~G.

The calculated angular integrals can also be used to study neutrino
emissivities in quark stars and hybrid stars or study cooling
properties of compact stars due to the emission of other weakly
interacting particles (for instance, axions; e.g.,
\citealt{Sedrak16}).

In a crust of a neutron star one can deal with neutrino reactions of
atomic nuclei and degenerate electrons (e.g.
\citealt{YKGH01,BK01,BK02}). For instance, it can be neutrino-pair
bremsstrahlung in electron-nucleus colli\-si\-ons or Urca cycles
involving Urca pairs of atomic nuclei. In these cases the nuclei do
not behave as strongly degenerate fermions and the neutrino
emissivities are not directly expressed through the angular
integrals $I_\Omega$ [although may contain similar integrals
$\widetilde{I}_\Omega$, Eq.\ (\ref{e:tI1})].

%%%%%%
\acknowledgements The authors are indebted to P. Shternin for strong
and constructive criticism and to K. Levenfish for encouragement.
The work by DY has been supported partly by the RFBR (grants
14-02-00868-a and 16-29-13009-ofi-m) and the work by PH by the
Polish NCN research grant no. 2013/11/B/ST9/04528. One of the
authors (A.D.K.) is grateful to N. Copernicus Astronomical Center in
Warsaw for hospitality and perfect working conditions.

%\smallskip

%\noindent
%\verb!\cite{bag02}! -- \cite{bag02}\\
%\verb!\citep{bag02}! -- \citep{bag02}\\
%\verb!\citet{bag02}! -- \citet{bag02}\\
%\verb!\cite{ale94}! -- \cite{ale94}\\
%\verb!\cite{bag02,ale94}! -- \cite{bag02,ale94}\\
%\verb!\citeauthor{ale94}! -- \citeauthor{ale94}\\
%\verb!\citeyear{ale94}! -- \citeyear{ale94}\\

%\nocite{*}
%%%  Using BibTeX  (Name-Year style)
%
\bibliographystyle{copernicus}

\end{document}